\documentclass[aps,preprint]{revtex4}
\usepackage[dvips]{epsfig}

\begin{document}


\title{Water adsorption on amorphous silica surfaces: A Car--Parrinello simulation study}
\author{Claus Mischler$^{1)}$, J\"urgen Horbach$^{1)}$, Walter Kob$^{2)}$, and Kurt Binder$^{1)}$}
\affiliation{$^{1)}$Institut f\"ur Physik, Johannes Gutenberg--Universit\"at Mainz, Staudinger Weg 7,
             D--55099 Mainz, Germany\\
             $^{2)}$Laboratoire des Collo\"ides, Verres et Nanomat\'eriaux, Universit\'e Montpellier II,
             Place E. Bataillon cc 069, F--34095 Montpellier, France}

%
\begin{abstract}
A combination of classical molecular dynamics (MD) and {\it ab
initio} Car--Parrinello molecular dynamics (CPMD) simulations is
used to investigate the adsorption of water on a free amorphous silica
surface. From the classical MD SiO$_2$ configurations with a free surface
are generated which are then used as starting configurations for the CPMD.
We study the reaction of a water molecule with a two--membered ring at the
temperature $T=300$~K.  We show that the result of this reaction is the
formation of two silanol groups on the surface. The activation energy of
the reaction is estimated and it is shown that the reaction is exothermic.
\end{abstract}

\maketitle


%
\section{Introduction}

The interaction of water with surfaces of amorphous silica is of
great technological interest~\cite{legrand98,iler79}, and thus
numerous studies have been devoted to this issue ranging from IR
spectroscopy~\cite{morrow76,michalske84,bunker89,dubois93a,dubois93b,grabbe95}
to numerical methods using both {\it ab initio}
techniques~\cite{keeffe84,uchino98,chuang96,masini02,du03a,du03b,du04}
and classical MD simulations~\cite{webb98,bakaev99,wilson00,walsh00}. In these
studies, the possible existence of small--membered Si--O rings on SiO$_2$
surfaces has been intensely discussed since such rings are expected to be
the reactive centres for the interaction with water and other molecules.
It is well--known that water may dissociate on SiO$_2$ surfaces resulting
in the formation of silanol (Si--OH) groups.  These silanol groups can
be detected in spectroscopic experiments via the Si--OH stretching
mode near 3750~cm$^{-1}$~\cite{hair,ryason75,morrow90}. In particular it
is believed that the silanol groups are a result of
the interaction of water molecules with small--membered
rings~\cite{morrow76,bunker89,dubois93a,dubois93b,grabbe95}.

In this paper we investigate the reaction of a water molecule with a
two--membered ring on an amorphous silica surface. To this end, we combine
classical MD simulations using the so--called BKS potential~\cite{bks90}
for silica and {\it ab initio} Car--Parrinello MD (CPMD)~\cite{car85}.
This combination of methods has recently also been used for investigating silicates 
in the bulk~\cite{benoit00,ispas01} and of free
silica surfaces~\cite{mischler02}.  The idea in our case is to generate
configurations with a free surface by ``BKS--MD'' which are then the
starting configurations for CPMD runs. The advantages of this methodology
are two--fold: On the one hand, in the classical MD amorphous systems can
be relaxed on a nanosecond time scale whereas typical time scales for the
CPMD are only of the order of several picoseconds.  On the other hand, the CPMD
provides a much more realistic modeling of atomistic systems. This is of
special importance in the case of interactions of water with an SiO$_2$
surface since it is difficult to model these interactions realistically
by classical potentials.

The rest of the paper is organized as follows: In the next section
we describe how we have prepared a free silica surface by 
classical MD and give the details of the CPMD simulations. 
Then, the results for the water interaction with the two--membered
ring are presented in Sec. 3. Finally, we summarize the results.

\section{Preparation of the surface and details of the simulations}

In order to study the reaction of a silica surface with a water molecule,
we first have to prepare SiO$_2$ configurations with a free surface.
As we have already mentioned in the Introduction we use for this a combination
of classical MD and CPMD.

As a model potential for the classical MD we use the so--called BKS
potential~\cite{bks90} which is a simple pair potential of the following
form:
\begin{equation}
  \phi(r)=
    \frac{q_{\alpha} q_{\beta} e^2}{r} +
     A_{\alpha \beta} \exp\left(-B_{\alpha \beta}r\right) -
    \frac{C_{\alpha \beta}}{r^6}\quad \alpha, \beta \in
    [{\rm Si}, {\rm O}] \ ,
    \label{bks}
\end{equation}
where $r$ is the distance between an atom of type $\alpha$ and an atom of
type $\beta$. The effective charges are $q_{{\rm O}} = -1.2$ and $q_{{ \rm
Si}} = 2.4$, and the parameters $A_{\alpha \beta}$, $B_{\alpha \beta}$,
and $C_{\alpha \beta}$ can be found in the original publication. They were
determined by using a combination of {\it ab initio} calculations and classical
lattice dynamics simulations.  The long--ranged Coulomb forces (and the
potential) were evaluated by means of the Ewald summation technique.
As an integrator for the simulation we used the velocity form of the
Verlet algorithm~\cite{allen} with a time step of $1.2$~fs.

In order to investigate a system with free surfaces one could consider a film
geometry, i.e.~a system with periodic boundary conditions (PBC) in two
directions and free boundaries in the third direction. Unfortunately,
this is not a very good approach since the Ewald summation technique for
the long ranged Coulomb interactions becomes inefficient in this case.
This stems from the fact that the Fourier part of the Ewald sums can no
longer be calculated by a single loop over the number of particles $N$
as in the case of PBC in all three directions, but one has to compute
a double loop that scales with $N^2$~\cite{parry75,parry76,leeuw82}.
Therefore, we have adopted in this work a different strategy in that
a system with PBC in all three directions was simulated containing an
empty space $\Delta z$ in $z$--direction.

The preparation of the interface is done by the following steps:  i)
We start with a system at $T=3400$~K with PBC in all three directions
(box dimensions: $L_x=L_y=11.51$~\AA~and $L_z^{\prime}=23$~\AA). This
system is fully equilibrated within 1~ns (see also below). As a result,
configurations are obtained that indicate a realistic modelling
of SiO$_2$ with the empirical BKS potential, Eq.~(\ref{bks}): A
tetrahedral Si--O network is formed containing defects (e.g.~five--fold
coordinated silicon atoms) which are expected at a temperature as high
as $T=3400$~K. However, no artificial bonds such as Si--Si or O--O bonds
are found in the resulting network structures. ii) We cut the system
perpendicular to the $z$--direction into two pieces. This is only
done for oxygen--silicon bonds such that we get only free oxygen atoms
at this interface. iii) These free oxygen atoms are now saturated by
hydrogen atoms. The place of these hydrogen atoms is chosen such that
each of the new oxygen--hydrogen bonds is in the same direction as the
oxygen--silicon bonds which were cut and have a length of approximately
$1$~\AA. The interaction between the hydrogen and the oxygen atoms as
well as the silicon atoms are described only by a Coulombic term. The
value of the effective charge of the hydrogen atoms is set to 0.6 which
ensures that the system is still (charge) neutral. iv) We permanently freeze atoms
which have a distance from the interface that is less than $4.5$~\AA,
whereas atoms that have a larger distance can propagate subject to the
force field, thus generating a mobile layer of 14.5~\AA. v) We add in
$z$--direction an empty space $\Delta z= 6.0$~\AA~. With this sandwich
geometry we can use periodic boundary conditions in all three directions.
We have made checked that the value of $\Delta z$ is sufficiently large
to ensure that the structure of the free surface is not affected by the
immobilized part of the system~\cite{mischler_diss}.  Finally, we have
a system of 91 oxygen, 43 silicon, and 10 hydrogen atoms in a simulation
box with $L_x=L_y=11.51$~\AA~and $L_z\approx 25$~\AA.

We have done 125 independent runs where we have equilibrated the system for
1 ns. For the present study we have selected from these runs one configuration
that has exactly one two--membered ring on
its surface. This configuration was then quenched to $T=300$~K using
a cooling rate of $2.8 \times 10^{12}$~K/s where it was further relaxed for
0.25~ns. Subsequently the final configuration of this run has been used as a starting point
for the CPMD. 

In the CPMD, we used conventional pseudopotentials for silicon and oxygen
and the BLYP exchange functions~\cite{trouiller,lee}. The electronic
wave--functions were expanded in a plane wave basis set with an energy
cutoff of 60~Ry and the equations of motion were integrated with a time
step of 0.085~fs for 0.2~ps. In the analysis of the CPMD run only those
configurations were taken into account that were produced later than
5~fs after the start of the CPMD run in order to allow the system to
equilibrate at least locally~\cite{benoit00}.

\section{Results}
\begin{figure}
\centering
\includegraphics[angle=0,width=0.8\linewidth]{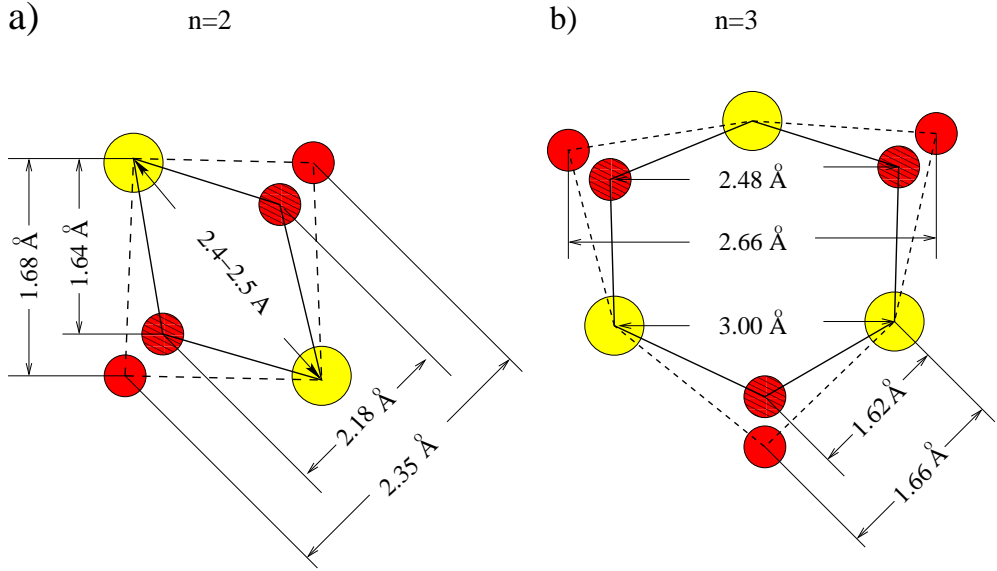}
\caption{Geometry of a) two--membered rings and b) three--membered 
         rings as obtained from BKS--MD (indicated by solid lines) 
         and from CPMD (indicated by the dashed lines). Si atoms 
         are shown as light circles and O atoms as small dark circles.
         For the sake of clarity of the figure, distances and angles
         are not to scale, and hence the difference between BKS--MD
         and CPMD in the figure is somewhat exaggerated in comparison with
         reality.}
  \label{fig1}
\end{figure}
On intermediate length scales the surface structure of SiO$_2$ can be
well distinguished from its bulk structure by the ring distribution~\cite{roder01}.
A ring is defined as the shortest closed loop of $n$ consecutive Si--O elements.
The SiO$_2$ surfaces that were obtained from the BKS--MD at $T=3400$~K
showed a relatively high concentration of 2--membered rings, i.e.~edge--sharing
tetrahedra, in contrast to the bulk in which, even at high temperatures, 
such structures are basically absent.
The relaxation of the surfaces by CPMD reduced
the number of 2--membered rings significantly at $T=3400$~K. At
$T=300$~K, the time scale on which the system is relaxed by means of CPMD
(0.2~ps) is not sufficient to cause a rearrangement of ring structures
and thus the number of 2--membered rings does not change during the
CPMD~\cite{mischler02,mischler_diss}.  However, both at $T=3400$~K and at
$T=300$~K, the CPMD yielded a slightly different geometry of 2--membered
rings which is illustrated in Fig.~\ref{fig1}a. For BKS--MD as well as CPMD one finds a
nearly planar geometry of two--membered rings.  Whereas in CPMD
a nearly quadratic shape with an O--Si--O angle around 90$^o$ is seen,
the BKS--MD leads to a trapezoid geometry with an O--Si--O angle around
80$^o$. This is due to the fact that the distance between nearest Si--O
and O--O neighbors is larger in CPMD while the distance between the
silicon atoms in the 2--fold ring is approximately around 2.5~\AA~in
both methods. A similar behavior can be seen for the 3--fold rings (see
Fig.~\ref{fig1}b) which exhibit also a nearly planar geometry. Also, in
this case BKS--MD and CPMD show a similar Si--Si distance (between nearest
Si neighbors) around 3.0~\AA~while the Si--O distance is slightly larger
in the CPMD. As a consequence, in the BKS--MD, the typical O--Si--O angle
is 99$^o$ whereas in the CPMD an O--Si--O of 107$^o$ is found which is
close to the angle of 109.47$^o$ in an ideal tetrahedron.

Small--membered rings are of particular interest if one considers the
reaction of a water molecule with an SiO$_2$ surface. A possible
reaction mechanism which was put forward by molecular orbital
calculations~\cite{keeffe84,uchino98,chuang96} implies the disruption of
a Si--O--Si structural element followed by the formation of two silanol
groups (Si--O--H). Thereby a Si--O bond has to be broken. For this reason,
2-- or 3--membered rings are of particular interest since these structures have a high local 
internal stress and hence their bonds can be broken more easily than 
the one in larger rings.  
Therefore, we consider in the following the reaction of H$_2$O
molecule with a 2--membered ring on a silica surface. To this end, we
have selected a configuration from our BKS--CPMD simulations at 300~K
that exhibit exactly one two--fold ring on its surface.

\begin{figure}[t]
  \vspace*{-1.4cm}
  \centering 
  \includegraphics[angle=0,width=0.6\linewidth]{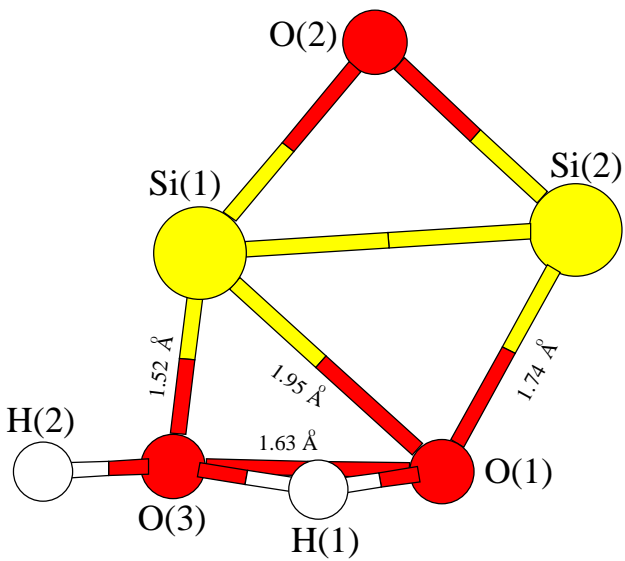}
  \vspace*{-2.7cm}
  \caption{Snapshot that illustrates the preparation of the water
           molecule on a two--fold ring. Note that Si(1), Si(2), O(1),
           and O(2) are part of the Si--O network.}
  \label{fig2}
\end{figure}
The initial position that we have chosen for the H$_2$O molecule on the
2--membered ring is illustrated in Fig.~\ref{fig2}. The oxygen atom of the
water molecule [O(3)] was placed at a distance of 1.52~\AA~from one of the
Si atoms [Si(1)] which is slightly smaller than the Si--O bond lengths
in the ring of 1.64~\AA. The two hydrogen atoms H(1) and H(2) were placed
at a distance of 1~\AA~from O(3) (and H(1) at a distance of 1~\AA~from
O(1)) such that the H(1)--O(3)--H(2) angle corresponds to 109$^{o}$.
Further details can be extracted from Fig.~\ref{fig2}. We shall see in
the following that the small O(1)--O(3) distance (1.63~\AA~compared
to 2.2~\AA~in bulk silica) drives the breaking of a Si--O bond and the
formation of two silanol groups.

\begin{figure}
\centering
  \includegraphics[angle=0,width=0.7\linewidth]{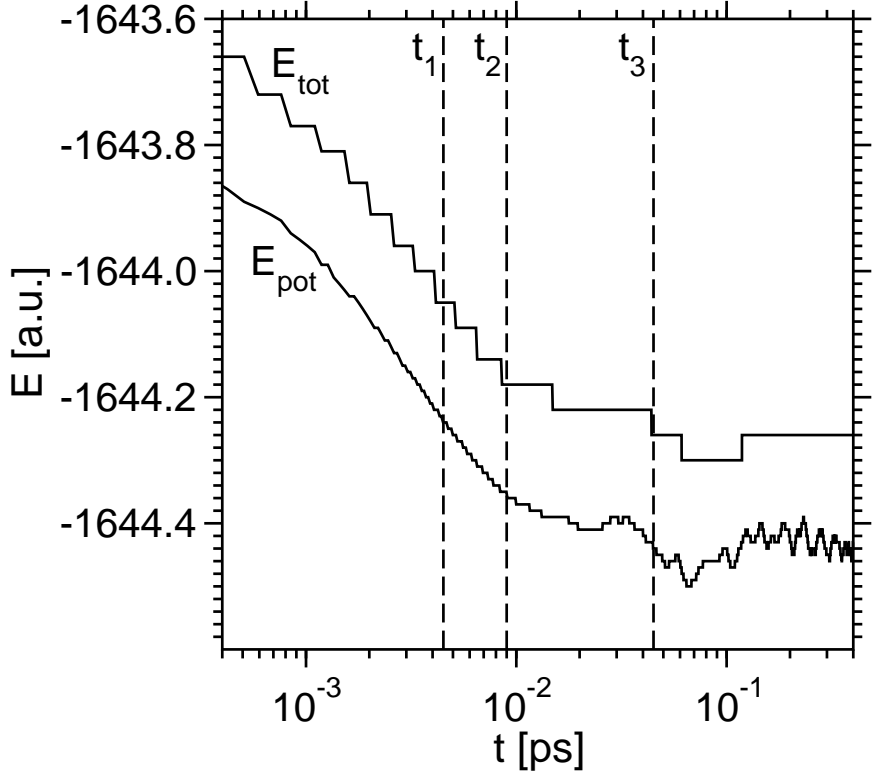}
  \caption{Total energy $E_{\rm tot}$ and potential energy $E_{\rm pot}$ 
           during the interaction of the silica surface with the H$_2$O 
           molecule (note the logarithmic time axis). The vertical dashed 
           lines at $t_1=0.0045$~ps, $t_2=0.009$~ps, and $t_3=0.045$~ps 
           mark the times at which the snapshots in Fig.~\ref{fig3abc} 
           are shown.}
  \label{fig3}
\end{figure}
\begin{figure}
\begin{center}
\begin{minipage}[t]{100mm}
\unitlength1mm
\begin{picture}(10,30)
\put(-67,-23){
\epsfysize=40mm
\epsffile{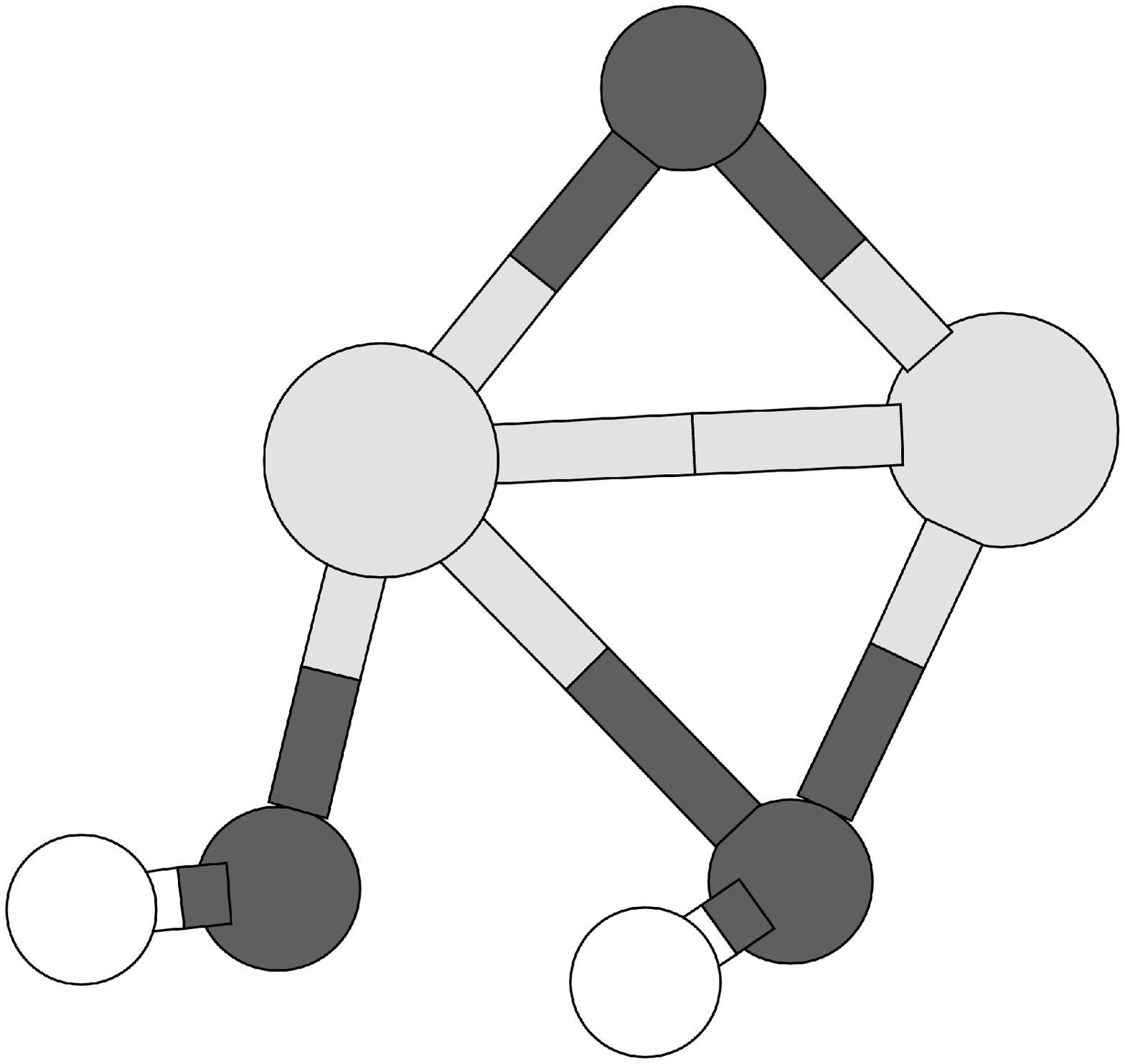}
}
\put(-67,18){
\mbox{\large{a)}}
}
\put(-15,-23){
\epsfysize=40mm
\epsffile{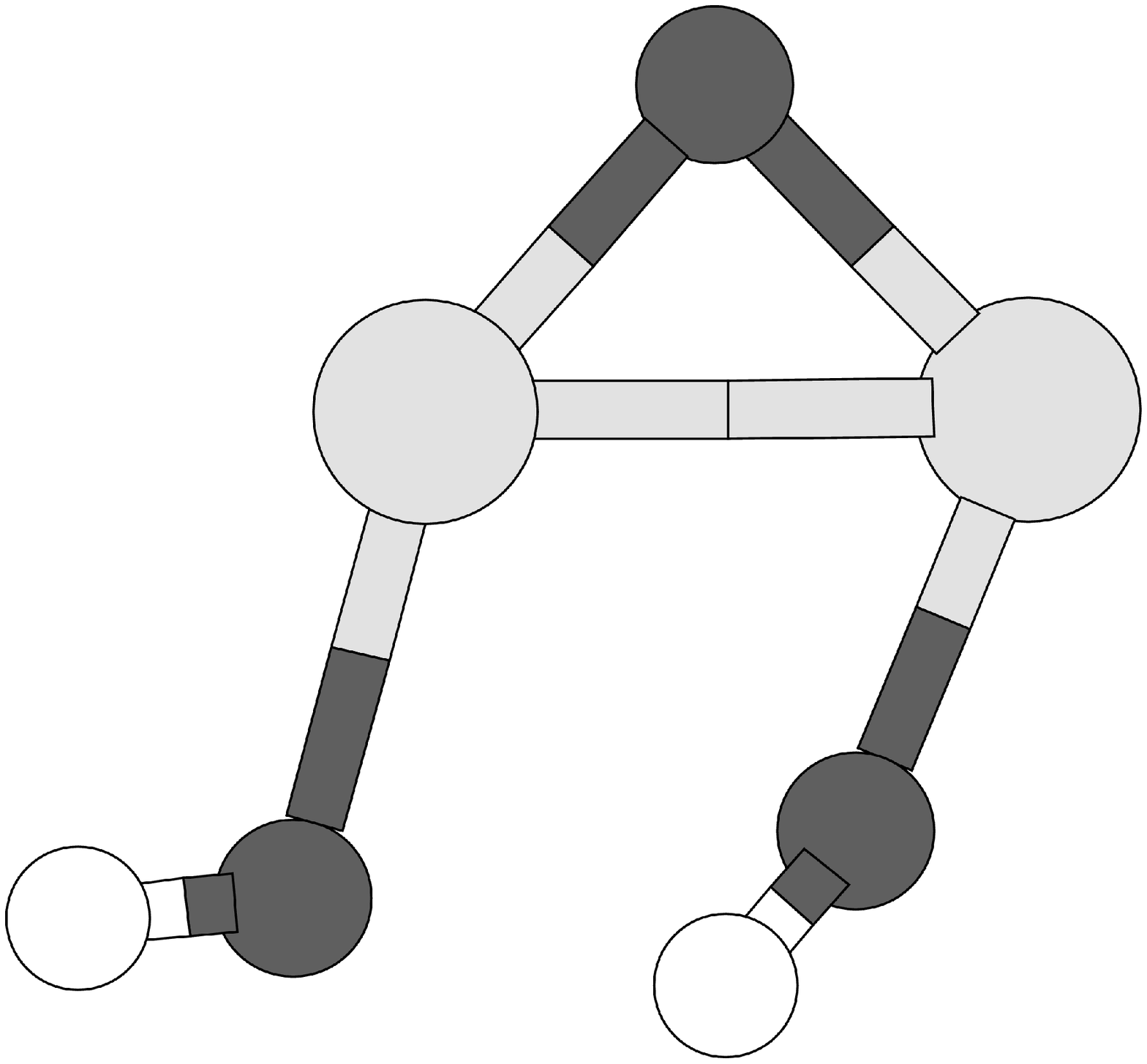}
}
\put(-15,18){
\mbox{\large{b)}}
}
\put(35,-23){
\epsfysize=35mm
\epsffile{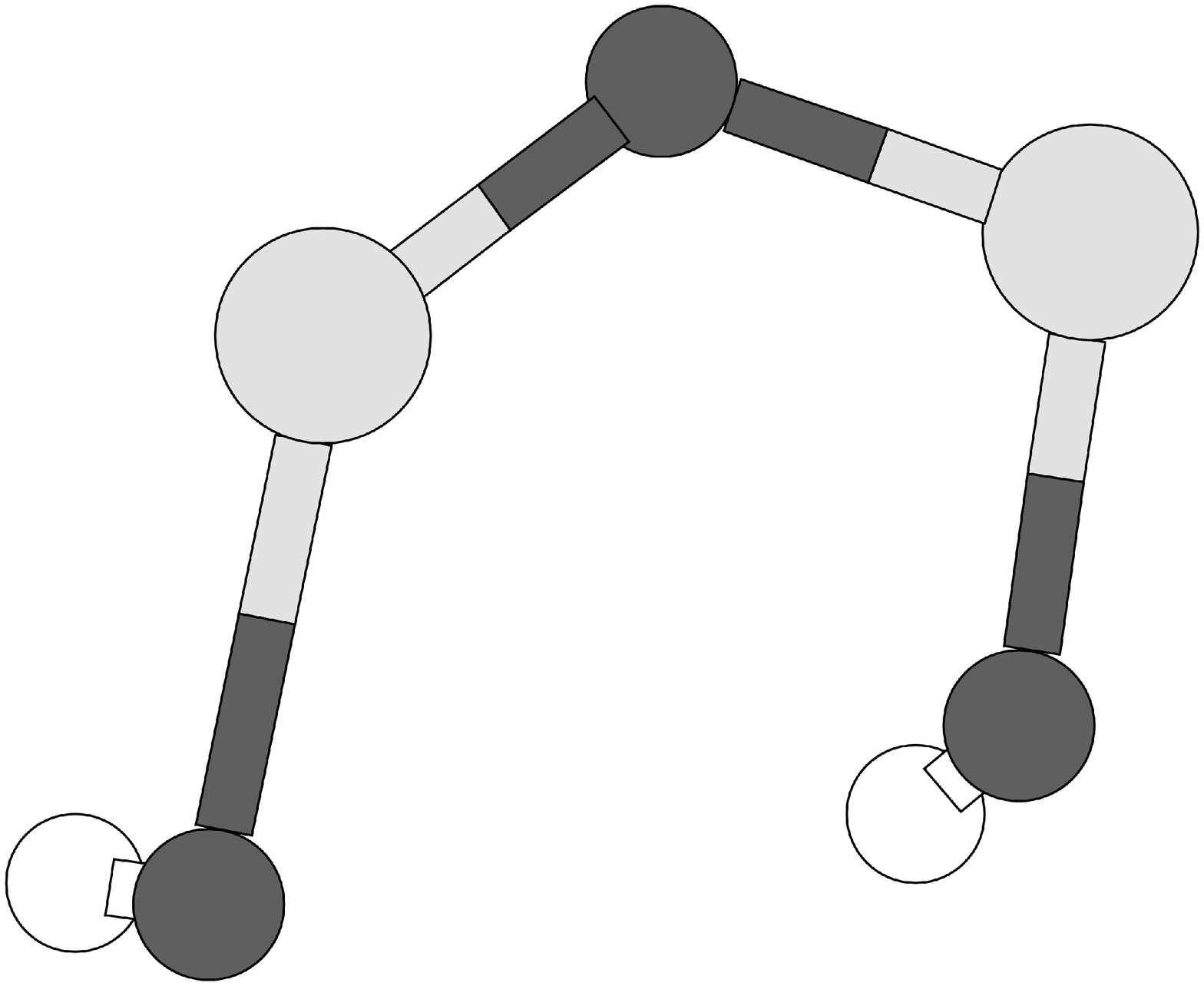}
}
\put(35,18){
\mbox{\large{c)}}
}
\end{picture}
\end{minipage}
\end{center}
\vspace{20mm}
\caption{Snapshots of the reaction of the H$_2$O molecule with 
the 2--fold ring at a) $t=t_1$, b) $t=t_2$, and c) $t=t_3$ 
($t_1$, $t_2$, and $t_3$ as indicated in Fig.~\ref{fig3}).}
\label{fig3abc}
\end{figure}
The initial condition for H$_2$O on the two--membered rings is of
course artificial and leads to enormous forces between the atoms. In
order to avoid an explosion of the system we have introduced a thermostat.
If the temperature
becomes larger than 380~K or lower than 220~K, the velocities of the
particles are rescaled such that the apparent temperature of the system
is 300~K.  Figure~\ref{fig3} shows the potential and the total energy
during the reaction whereby the time $t=0$ corresponds to the initial
configuration. Obviously, the potential energy decreases rapidly such
that after about 0.1~ps a constant is reached.  Snapshots of the reaction
at $t_1=0.0045$~ps, $t_2=0.009$~ps, and $t_3=0.045$~ps are shown in
Fig.~\ref{fig3abc} (note that $t_1$, $t_2$, and $t_3$ are also marked
as vertical dashed lines in Fig.~\ref{fig3}). Indeed, at time $t_3$
the end configuration with two silanol groups is formed. The first step
that leads to the silanol groups is the breaking of one of the H--O
bonds and of the enforced O(3)--O(1) bond. This has happened at $t=t_1$
whereby the potential energy is already quite close to the equilibrium value
(see Fig.~\ref{fig3}). In a second step then a Si--O bond is broken
(see snapshot at $t=t_2$) followed by an increased separation of the two
silicon atoms.

\begin{figure}[t]
\vspace*{-1.5cm}
\begin{center}
\begin{minipage}{110mm}
\unitlength1mm
\begin{picture}(10,10)
\put(-55,-82){
\epsfysize=75mm
\epsffile{fig_deltar.eps}
}
\put(30,-54){
\epsfysize=53mm
\epsffile{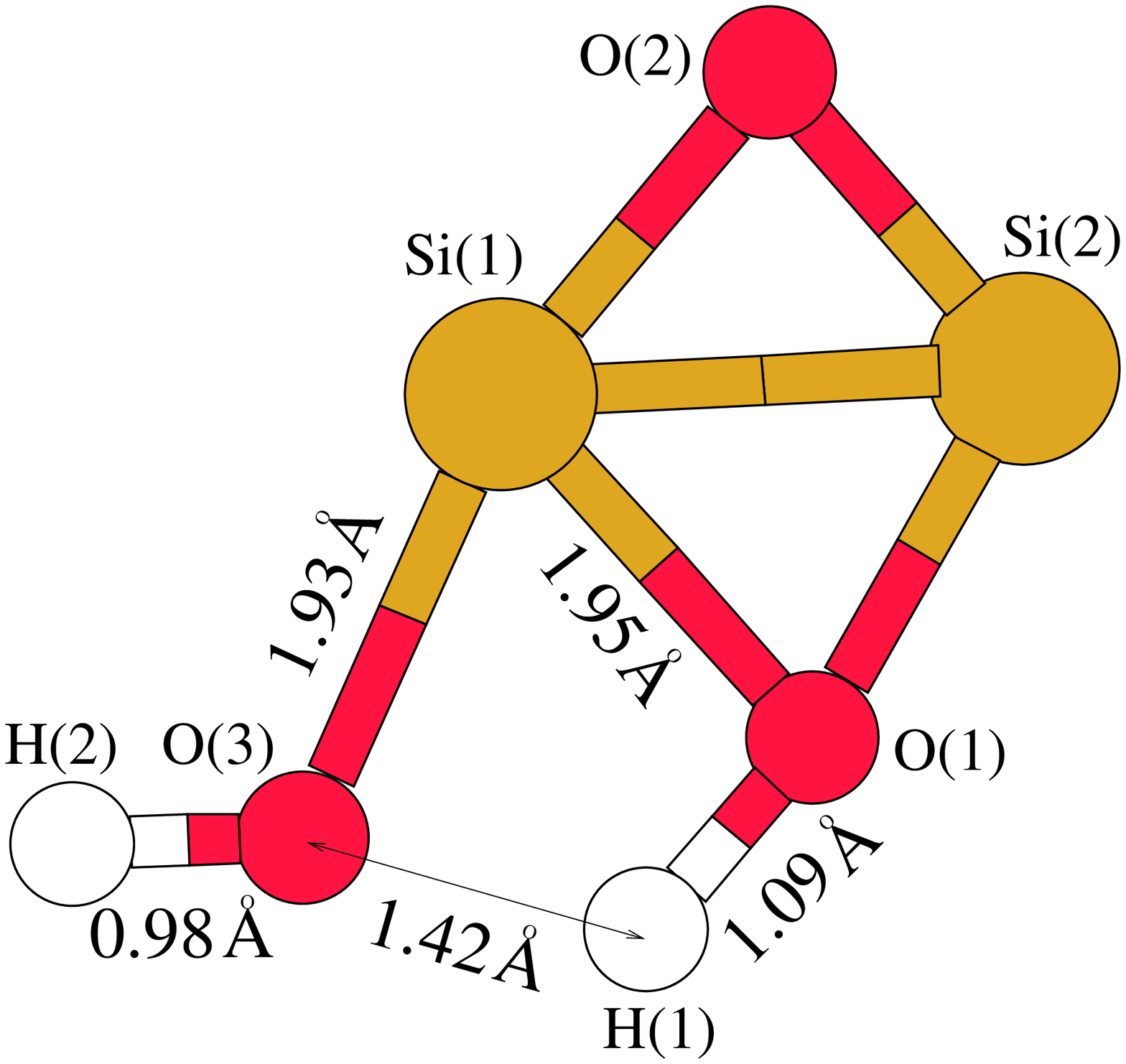}
}
\put(30,-90){
\epsfysize=53mm
\epsffile{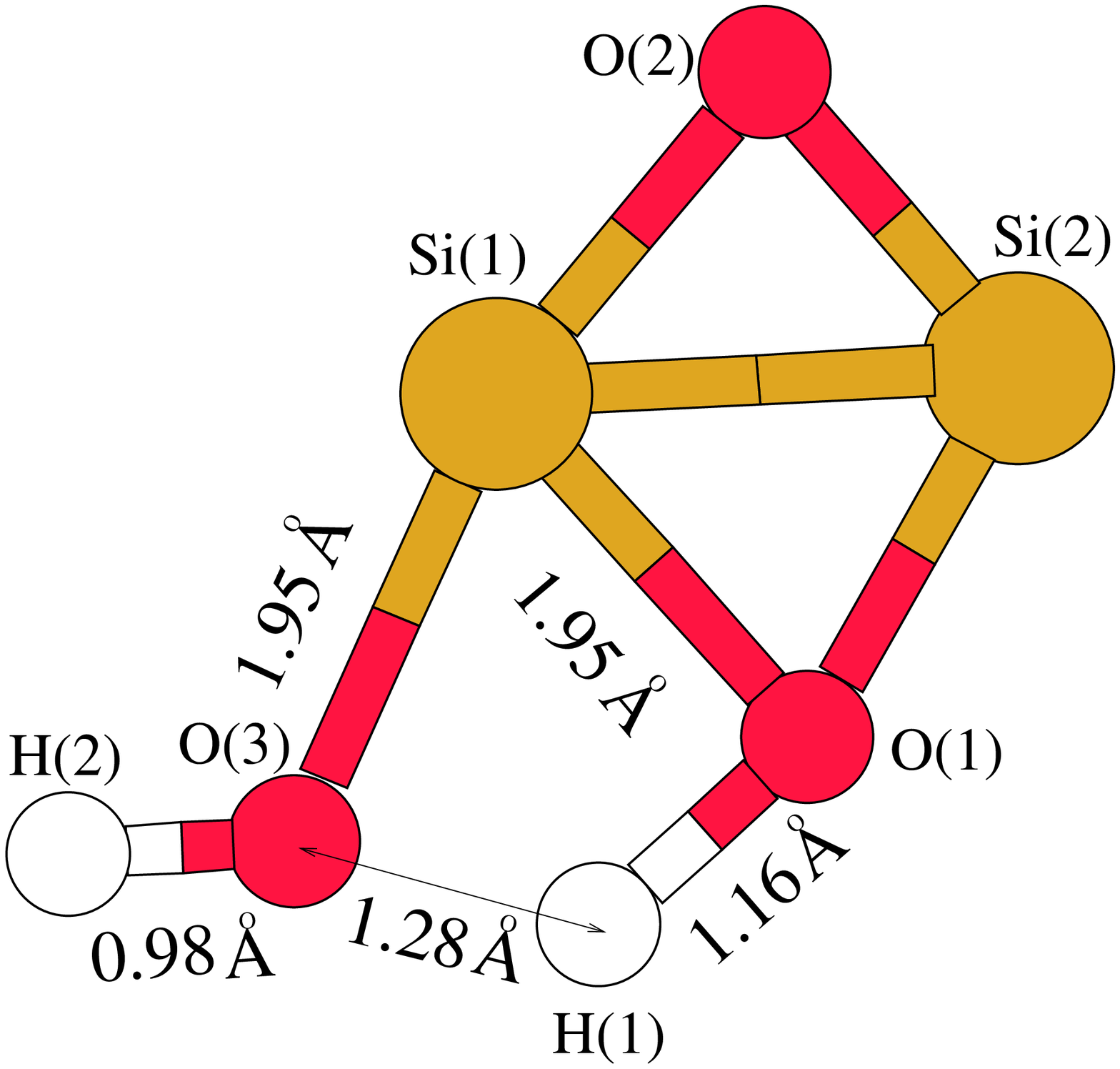}
}
\end{picture}
\end{minipage}
\end{center}
\vspace{80mm}
\caption{Change in the distance $\delta r$ between the indicated atoms
         (that are defined in the two snapshots at times $t_1^{\rm r}=0.044$~ps
         and $t_2^{\rm r}=0.047$~ps) during the simulation at $T=30$~K where
         only the H$_2$O molecule was moved whereas the other atoms
         were fixed. The inset shows the potential energy for the latter
         simulation (note that $E_{\rm pot}$ is shifted by 1640~a.u.).}
\label{fig_deltar}
\end{figure}

Now we want to study in more detail the formation of two silanol
groups as a result of the reaction of the water molecule with a
two--fold ring. This reaction can be written as follows:
\begin{equation}
\begin{picture}(0,100)
\put(-150,5){
\epsfysize=30mm
\epsffile{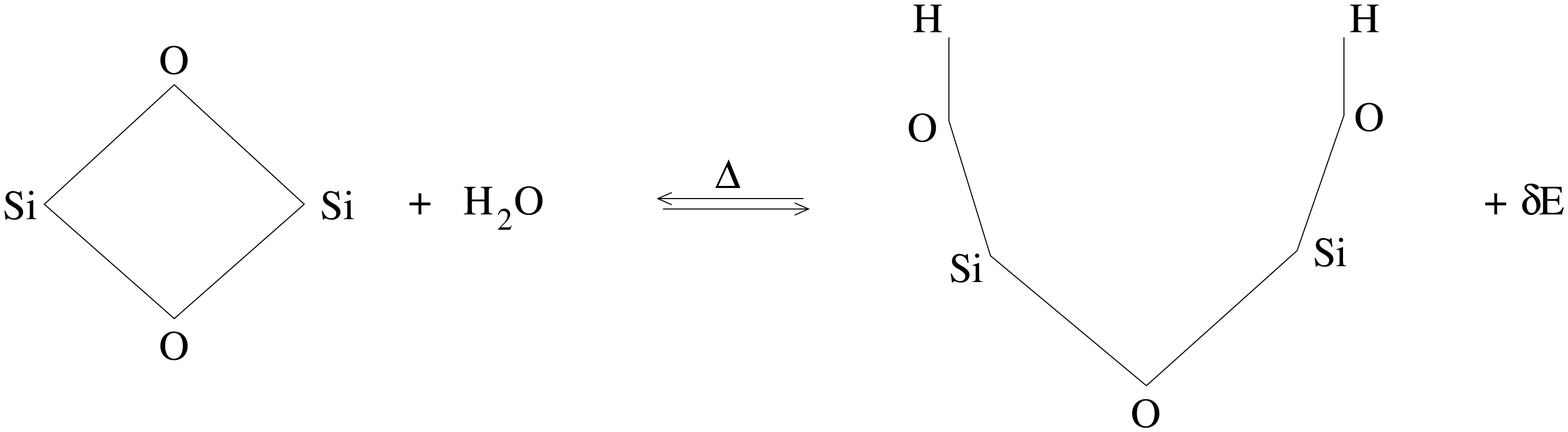}
}
\end{picture}
\label{sio2h2o}
\end{equation}
\begin{figure}[t]
\vspace*{-3.7cm}
\begin{center}
\begin{minipage}{110mm}
\unitlength1mm
\begin{picture}(10,30)
\put(-60,-78){
\epsfysize=70mm
\epsffile{fig5.eps}
}
\put(30,-34){
\epsfysize=25mm
\epsffile{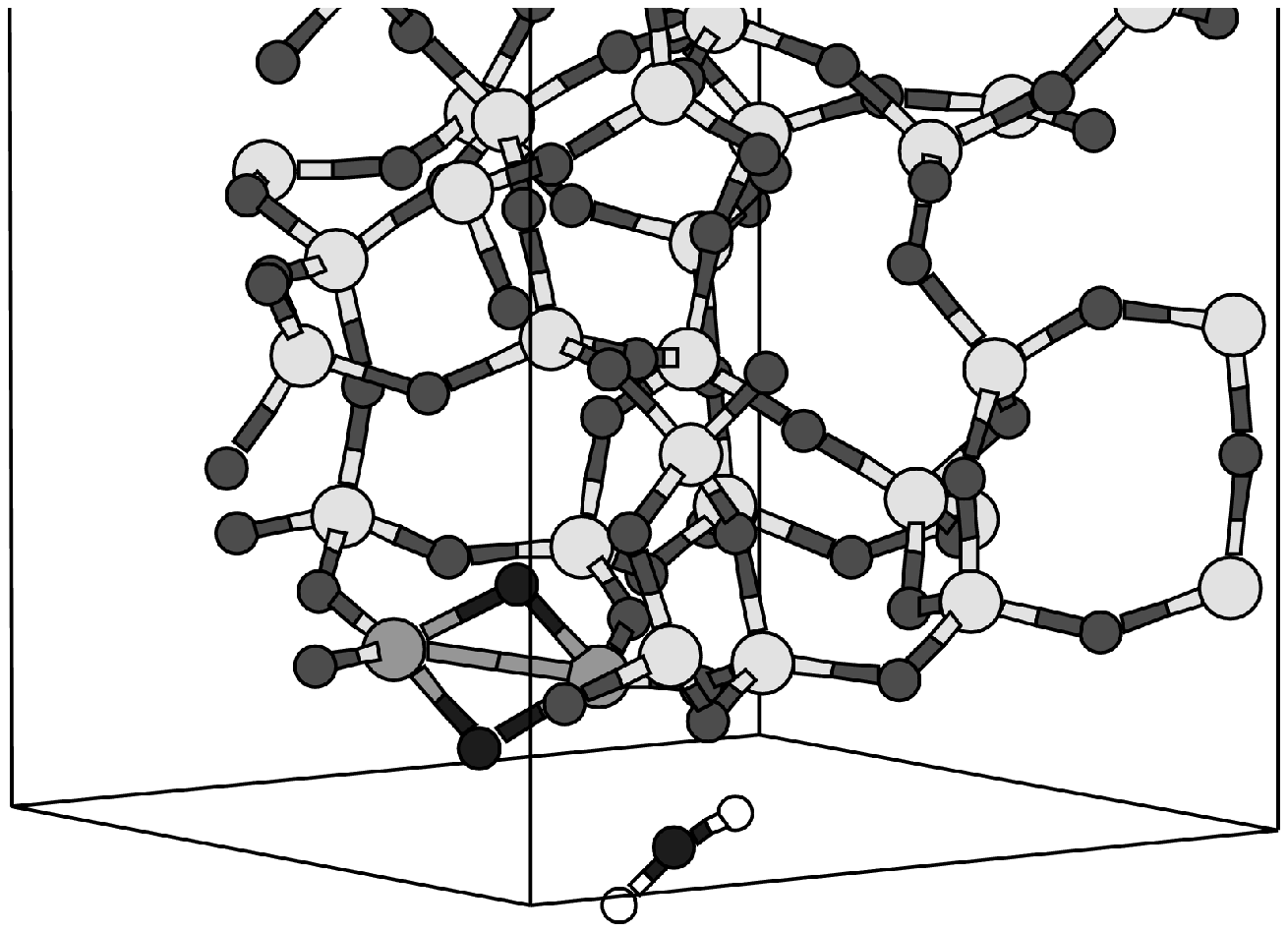}
}
\put(30,-68){
\epsfysize=25mm
\epsffile{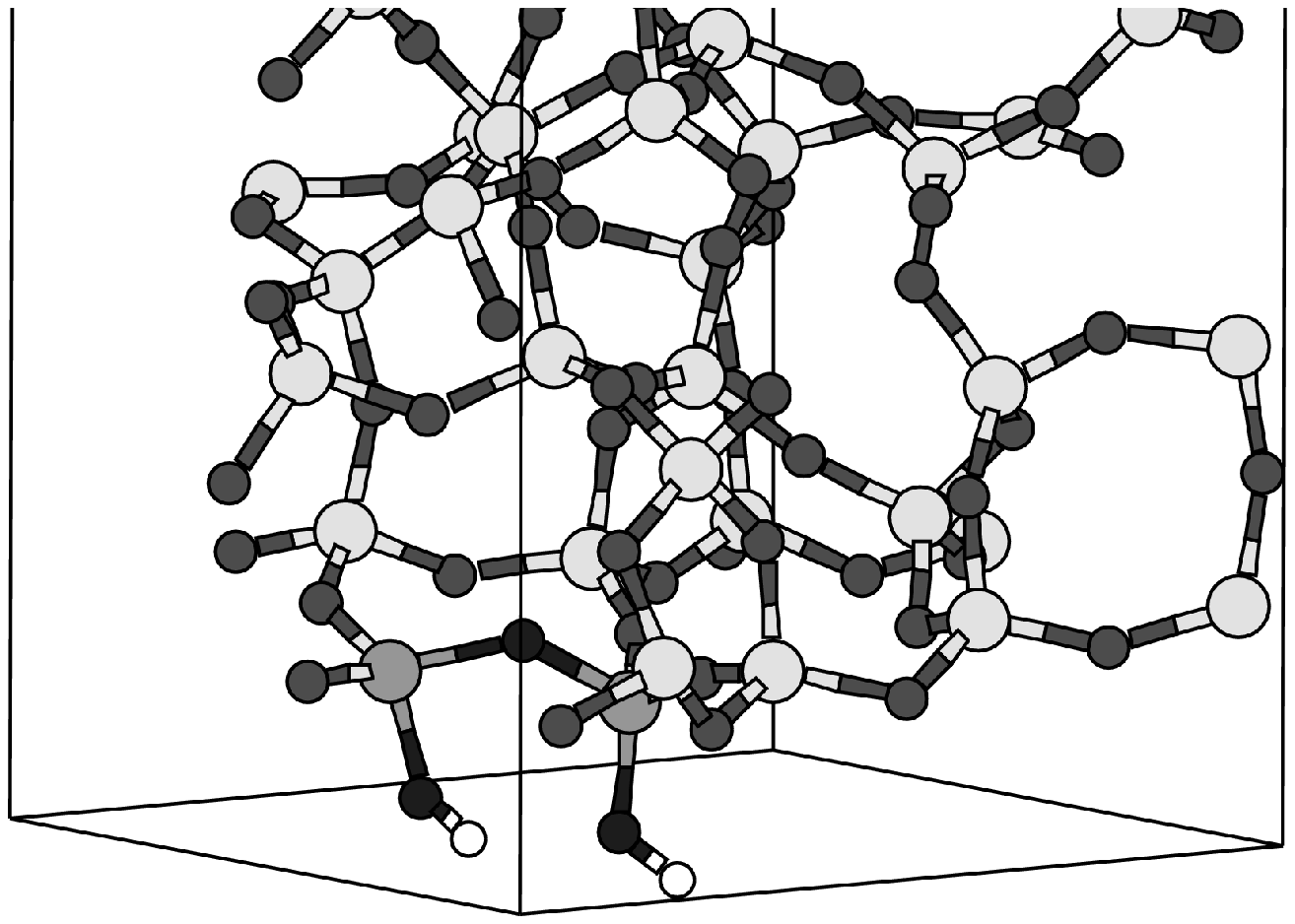}
}
\end{picture}
\end{minipage}
\end{center}
\vspace{78mm}
\caption{Potential energy as a function of time for the simulations
         at $T=300$~K using the start configuration SC2
         (dashed line) and SC1 (solid line). The snapshots correspond
         to the end configurations of the two runs as indicated. In both 
         figures the two hydrogen atoms are shown as small white spheres at the 
         bottom of the simulation box.}
\label{fig5}
\end{figure}
In the following we will estimate the activation energy $\Delta$ and
also $\delta E$. A positive or negative sign of $\delta E$ indicates
whether the reaction is exothermic or endothermic, respectively.
We use again the starting configuration shown in Fig.~\ref{fig2}.
But now we completely freeze in the SiO$_2$ system and we move
only the water molecule at a temperature of $30$~K.  As the inset of
Fig.~\ref{fig_deltar} shows, the potential energy is monotonously decaying
while the water molecule relaxes on the surface. After about 0.1~ps a
constant value of $E_{\rm pot}$ is reached where the water molecule has
evaporated from the surface.  Also shown in Fig.~\ref{fig_deltar} are the
distances $\delta r$ of the water atoms from the initial position as a function of
time.  We see that, slightly before the H$_2$O evaporates,
the O(3)--O(1) distance and the O(3)--Si(1) distance are increasing
while the O(3)--H(1) distance is decreasing to the ideal O--H distance
in a water molecule around 1~\AA.  Of importance for the following,
are the configurations at the times $t_1^{\rm r}=0.044$~ps and $t_2^{\rm
r}=0.047$~ps (marked as vertical dashed lines in Fig.~\ref{fig_deltar})
that we will denote by SC1 and SC2, respectively. Using SC1 as a
starting configuration for a CPMD run, where the whole system is again
simulated at $T=300$~K, leads to the formation of two silanol groups,
whereas the use of SC2 yields the evaporation of the water molecule from
the surface. As we can infer from the snapshots in Fig.~\ref{fig_deltar},
the main difference between SC1 and SC2 is a slightly different O(3)--H(1)
distance.

In Fig.~\ref{fig5} we plot the potential energy for the two CPMD
runs starting from the configurations SC1 and SC2. Since both runs
yield initially about the same value for $E_{\rm pot}$ we can extract
both $\Delta$ and $\delta E$ from the plot. First, from the difference
of the saturated values for $t>0.1$~ps (marked as horizontal lines in
Fig.~\ref{fig5}) we obtain $\delta E = E({\rm SiO}_2 + {\rm H}_2{\rm O})
- E(2{\rm SiOH}) \approx 1.6$~eV which indicates that the reaction is
exothermic. The activation energy $\Delta$ can be estimated from the
difference of the initial and the final value of the SC2 run where the
water molecule evaporates from the surface. For the latter process an
HO--H bond and a Si--O bond have to be broken and thus the binding
energy of both bonds contribute to $\Delta$.  From Fig.~\ref{fig5}
the value $\Delta \approx 0.9$~eV can be extracted.  Whereas the
distances of the O--O and Si--O neighbors in the two--membered ring are
significantly smaller than those found in the bulk (see above), the same
bonds are very close to the bulk values in the final structure with the
two silanol rings.  Moreover, the O--Si--O angle in the silanol groups
has a value around 108$^o$, i.e.~close to the value found in an ideal
SiO$_4$ tetrahedron. These findings may explain the observation of
an exothermic reaction.

In a recent publication, Masini and Bernasconi~\cite{masini02}
also studied the adsorption reaction of a water molecule with a
two--membered ring on an amorphous silica surface by CPMD. They find
$\delta E \approx 1.7$~eV, i.e.~a value which is very similar to the
one found in this work. However, they did a constrained MD simulation
to estimate the activation energy.  As a reaction coordinate they used
either the distance between a silicon atom in the ring and the oxygen
atom of the water molecule (called path A in Ref.~\cite{masini02}) or
the distance between an oxygen atom of the ring and a hydrogen atom of
the water molecule (called path B). As a result $\Delta = 1.1$~eV is
obtained for path A and $\Delta = 0.32$~eV for path B.  In this work,
we find an activation energy which is closer to path A in 
Ref.~\cite{masini02}.

\section{Summary}

Using a combination of classical MD and CPMD we have investigated the
reaction of water with a free amorphous silica surface. We have seen that
the reaction of a water molecule with a two--membered ring, leading to the
formation of two silanol groups on the SiO$_2$ surface, is an exothermic
reaction with an activation energy around 0.9~eV.  That the reaction is
exothermic is in agreement with other recent numerical studies, and may
be explained by the change of bond lengths and angles towards values
found in bulk SiO$_2$ while the silanol groups form.

\section*{Acknowledgements}
We are grateful to Michele Parrinello and Gloria Tabacchi for the
introduction of one of us (C. M.) into CPMD.  Without their support the
present work would not have been possible.  We acknowledge financial
support by the SCHOTT Glaswerke Fond, the DFG under SFB 262, the BMBF
under grant No. 03N6015, and the European Community's Human Potential
Program under contract HPRN-CT-2002-00307, DYGLAGEMEM.  J.~H.~acknowledges
financial support from the DFG under grants HO 2231/2-1/2.  We thank
the NIC J\"ulich for a generous grant of computing time.


\begin{thebibliography}{10}
\expandafter\ifx\csname url\endcsname\relax
  \def\url#1{\texttt{#1}}\fi
\expandafter\ifx\csname urlprefix\endcsname\relax\def\urlprefix{URL }\fi

\bibitem{legrand98}
A. P. Legrand, {\it The Surface Properties of Silica}, Wiley, New York, 1998.

\bibitem{iler79}
R. K. Iler, {\it The Chemistry of Silica}, Wiley, New York, 1979.

\bibitem{morrow76}
B. A. Morrow, I. A. Cody, {\it J. Phys. Chem.}, {\bf 1976}, {\it 80}, 1995.

\bibitem{michalske84}
T. A. Michalske, B. C. Bunker, {\it J. Appl. Phys.}, {\bf 1984}, {\it 56}, 2686.

\bibitem{bunker89}
B. C. Bunker, D. M. Haaland, K. J. Ward, T. A. Michalske, W. L. Smith, J. S. Binkley,
C. F. Melius, C. A. Balfe, {\it Surf. Sci.}, {\bf 1989}, {\it 210}, 406.

\bibitem{dubois93a}
L. H. Dubois, B. R. Zegarski, {\it J. Am. Chem. Soc.}, {\bf 1993}, {\it 115}, 1190.

\bibitem{dubois93b}
L. H. Dubois, B. R. Zegarski, {\it J. Phys. Chem.}, {\bf 1993}, {\it 97}, 1665.

\bibitem{grabbe95}
A. Grabbe, T. A. Michalske, W. L. Smith, {\it J. Phys. Chem.}, 
{\bf 1995}, {\it 99}, 4648.

\bibitem{keeffe84}
M. O'Keeffe, G. V. Gibbs, {\it J. Chem. Phys.}, {\bf 1984}, {\it 81}, 876.

\bibitem{uchino98}
T. Uchino, Y. Tokuda, T. Yoko, {\it Phys. Rev. B}, {\bf 1998}, {\it 58}, 5322.

\bibitem{chuang96}
I. S. Chuang, G. E. Maciel, {\it J. Am. Chem. Soc.}, {\bf 1996}, {\it 118}, 401.

\bibitem{masini02}
P. Masini, M. Bernasconi, {\it J. Phys.: Condens. Matter}, {\bf 2002}, {\it 14}, 4133.

\bibitem{du03a}
M. H. Du, L. L. Wang, A. Kolchin, H. P. Cheng, {\it Eur. Phys. J. D}, {\bf 2003}, {\it 24}, 323.

\bibitem{du03b}
M. H. Du, A. Kolchin, H. P. Cheng, {\it J. Chem. Phys.}, {\bf 2003}, {\it 119}, 6418.

\bibitem{du04}
M. H. Du, A. Kolchin, H. P. Cheng, {\it J. Chem. Phys.}, {\bf 2004}, {\it 120}, 1044.

\bibitem{webb98}
E. B. Webb, S. H. Garofalini, {\it J. Non--Cryst. Sol.}, {\bf 1998}, {\it 226}, 47.

\bibitem{bakaev99}
V. A. Bakaev, W. A. Steele, {\it J. Chem. Phys.}, {\bf 1999}, {\it 111}, 9803.

\bibitem{wilson00}
M. Wilson, T. R. Walsh, {\it J. Chem. Phys.}, {\bf 2000}, {\it 113}, 9180.

\bibitem{walsh00}
T. R. Walsh, M. Wilson, A. P. Sutton, {\it J. Chem. Phys.}, {\bf 2000}, {\it 113}, 9191.

\bibitem{hair}
M. L. Hair, {\it Infrared Spectroscopy in Surface Chemistry}, Dekker, New York, 1967.

\bibitem{ryason75}
P. R. Ryason, B. G. Russell, {\it J. Phys. Chem.}, {\bf 1975}, {\it 79}, 1276.

\bibitem{morrow90}
B. A. Morrow, A. J. McFarlan, {\it J. Non--Cryst. Sol.}, {\bf 1990}, {\it 120}, 61.

\bibitem{bks90}
B. W. H. van Beest, G. J. Kramer, R. A. van Santen,
{\it Phys. Rev. Lett.} {\bf 1990}, {\it 64}, 1955.

\bibitem{car85}
R. Car, M. Parrinello, {\it Phys. Rev. Lett.} {\bf 1985}, {\it 55}, 2471.

\bibitem{benoit00}
M. Benoit, S. Ispas, P. Jund, R. Jullien,
{\it Eur. Phys. J. B} {\bf 2000}, {\it 13}, 631.

\bibitem{ispas01}
S. Ispas, M. Benoit, P. Jund, R. Jullien, 
{\it Phys. Rev. B}, {\bf 2001}, {\it 64}, 214206.

\bibitem{mischler02}
C. Mischler, W. Kob, K. Binder, {\it Comp. Phys. Comm.}, {\bf 2002}, {\it 147}, 222.

\bibitem{allen}
M. P. Allen, D. Tildesley, {\it Computer Simulation of Liquids}, Clarendon Press, Oxford, 1987.

\bibitem{parry75}
D. E. Parry,
{\it Surf. Sci.}, {\bf 1975}, {\it 49}, 433.

\bibitem{parry76}
D. E. Parry,
{\it Surf. Sci.}, {\bf 1976}, {\it 54}, 195.

\bibitem{leeuw82}
S. W. de Leeuw and J. W. Perram,
{\it Physica A}, {\bf 1982}, {\it 113A}, 546.

\bibitem{mischler_diss}
C. Mischler, 
{\it Molekulardynamik--Simulation zur Struktur von SiO$_2$--Oberfl\"achen
mit adsorbiertem Wasser}, Ph.D. Thesis, Mainz, 2002.

\bibitem{trouiller}
N. Trouiller, J. L. Martins, {\it Phys. Rev. B}, {\bf 1991}, {\it 43}, 1993.

\bibitem{lee}
C. Lee, W. Yang, R. G. Parr, {\it Phys. Rev. B}, {\bf 1988}, {\it 37}, 785.

\bibitem{roder01}
A. Roder, W. Kob, K. Binder,
{\it J. Chem. Phys.}, {\bf 2001}, {\it 114}, 7602.


\end{thebibliography}
\end{document}